\documentclass[aps, prl,reprint,amsmath,showpacs,amssymb,superscriptaddress]{revtex4-1}

\usepackage{graphicx}
\usepackage{dcolumn}
\usepackage{bm}

\begin{document}
\title{The Impact of Small-Angle Scattering on Ballistic Transport in Quantum Dots}

\author{A.M. See}
\affiliation{School of Physics, University of New South Wales,
Sydney NSW 2052, Australia}

\author{I. Pilgrim}
\affiliation{Department of Physics, University of Oregon, Eugene, OR
97403-1274, USA}

\author{B.C. Scannell}
\affiliation{Department of Physics, University of Oregon, Eugene, OR
97403-1274, USA}

\author{R.D. Montgomery}
\affiliation{Department of Physics, University of Oregon, Eugene, OR
97403-1274, USA}

\author{O. Klochan}
\affiliation{School of Physics, University of New South Wales,
Sydney NSW 2052, Australia}

\author{A.M. Burke}
\affiliation{School of Physics, University of New South Wales,
Sydney NSW 2052, Australia}

\author{M. Aagesen}
\affiliation{Nanoscience Center, Niels Bohr Institute, University of
Copenhagen, Universitetsparken 5, DK-2100 Copenhagen, Denmark}

\author{P.E. Lindelof}
\affiliation{Nanoscience Center, Niels Bohr Institute, University of
Copenhagen, Universitetsparken 5, DK-2100 Copenhagen, Denmark}

\author{I. Farrer}
\affiliation{Department of Physics, Cavendish Laboratory, J. J.
Thompson Avenue, Cambridge, CB3 0HE, United Kingdom}

\author{D.A. Ritchie}
\affiliation{Department of Physics, Cavendish Laboratory, J. J.
Thompson Avenue, Cambridge, CB3 0HE, United Kingdom}

\author{R.P. Taylor}
\affiliation{Department of Physics, University of Oregon, Eugene, OR
97403-1274, USA}

\author{A.R. Hamilton}
\affiliation{School of Physics, University of New South Wales,
Sydney NSW 2052, Australia}

\author{A.P. Micolich}
\affiliation{School of Physics, University of New South Wales,
Sydney NSW 2052, Australia}
\email{adam.micolich@nanoelectronics.physics.unsw.edu.au}

\date{\today}

\begin{abstract}

Disorder increasingly affects performance as electronic devices are
reduced in size. The ionized dopants used to populate a device with
electrons are particularly problematic, leading to unpredictable
changes in the behavior of devices such as quantum dots each time
they are cooled for use. We show that a quantum dot can be used as a
highly sensitive probe of changes in disorder potential, and that by
removing the ionized dopants and populating the dot
electrostatically, its electronic properties become reproducible
with high fidelity after thermal cycling to room temperature. Our
work demonstrates that the disorder potential has a significant,
perhaps even dominant, influence on the electron dynamics, with
important implications for `ballistic' transport in quantum dots.

\end{abstract}

\maketitle

Advances in semiconductor device technologies have enabled a long
and fruitful study of nanoscale devices obtained by further
confinement of the two-dimensional electron gas (2DEG) formed in an
AlGaAs/GaAs heterostructure~\cite{FerryBook09}. An important topic
is ballistic transport effects, which are traditionally considered
to occur when the large-angle scattering length exceeds the scale of
additional confinement~\cite{BeenakkerSSP91}. Following an early
focus on fundamental phenomena such as the Aharonov-Bohm
effect~\cite{TimpPRL87} and 1D conductance
quantization~\cite{vanWeesPRL88, WharamJPC88}, the potential for
novel devices was also explored~\cite{GoodnickIEEE03}. A highlight
with broad implications was the study of quantum chaos, where
quantum dots were used as model dynamical systems called
`billiards'~\cite{JalabertPRL90, MarcusPRL92, ChangPRL94,
JensenNat95}, alongside microwave~\cite{StockmannPRL90},
optical~\cite{GmachlSci98}, acoustic~\cite{SchaadtPRE03} and cold
atom systems~\cite{ZhangPRL04}. The physics of wave chaos should be
universal; however the various practical implementations differ,
with important consequences for observed
behavior~\cite{BerggrenChaos96}, as we demonstrate for semiconductor
billiards.

On its own, the large-angle scattering length $\ell$, measured via
the electron mobility, gives an incomplete picture of the overall
electron scattering in an AlGaAs/GaAs
heterostructure~\cite{ColeridgePRB91}. The 2DEG is normally
populated by ionization of Si dopants, with high mobility obtained
by spatially separating these dopants from the 2DEG. This
`modulation doping'~\cite{DingleAPL78} technique works because
increases in dopant-2DEG separation convert the ionized dopants from
large-angle to small-angle scattering sites that are effectively
`hidden' because the mobility is weighted towards large-angle
scattering~\cite{ColeridgePRB91}. Nonetheless, the 2DEG still
`feels' the ionized dopants as a low-level `disorder potential' with
a length scale set by the 2DEG-donor separation~\cite{NixonPRB90,
ColeridgePRB91, JuraNP07}. This small-angle scattering length scale,
typically of order $20 - 100$~nm, is much smaller than both the
typical quantum dot width ($\sim 0.6 - 2~\mu$m) and the large-angle
scattering length ($\sim 2 - 20~\mu$m). Although small-angle ionized
dopant scattering was considered in early studies of quantum dot
chaos~\cite{MarcusPRL92, LinChaos93}, it was generally expected to
have little influence on transport. Scanning gate microscopy (SGM)
allows direct visualization of electron flow in nanoscale
devices~\cite{TopinkaNat01, JuraNP07, AidalaNP07}, and small-angle
scattering clearly causes significant deviations from the straight
trajectories envisioned in the ballistic transport
paradigm~\cite{BeenakkerSSP91, JalabertPRL90, JensenNat95}. This
raises two important questions: What is the true impact of
small-angle scattering on transport, as measured by the conductance,
in quantum dots? Can its effect be reduced or eliminated?

\begin{figure}
\includegraphics[width = 8 cm]{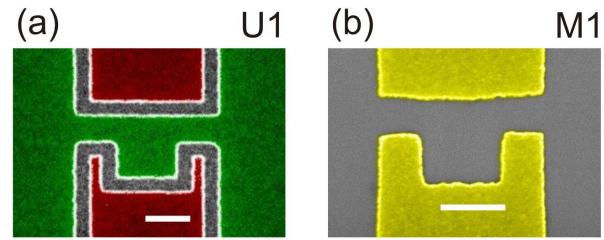}
\caption{(a/b) Scanning electron micrographs of undoped device U1
and modulation-doped device M1 with $500$~nm scale bars (white).
Green and red tinted regions in (a) indicate the n$^{+}$ GaAs top-
and side-gates. Yellow tinted regions in (b) are Ti/Au gates on the
heterostructure surface.}
\end{figure}

We address both questions by measuring the low temperature
magnetoconductance $G(B)$ of two quantum dots (Fig.~1) with
nominally identical geometry, one on undoped (U1) and one on
modulation-doped (M1) heterostructure, before and after thermal
cycling to room temperature. Our undoped dot design evolved from the
Heterostructure Insulated Gate Field-Effect Transistor (HIGFET)
conceived by Solomon {\it et al}~\cite{SolomonEDL84}. The
heterostructure consists of an undoped GaAs substrate overgrown with
$160$~nm undoped AlGaAs, $25$~nm undoped GaAs, and a $35$~nm n$^{+}$
GaAs cap. The cap is highly conductive at low temperature and
divided into three independently biasable gates (Fig.~1(a)). A
positive bias $V_{TG} > 0.32$~V applied to the top-gate (green)
electrostatically populates the dot and source and drain reservoirs.
A negative voltage $V_{SG}$ applied to the side-gates (red) tunes
the dot area and width of the quantum point contacts (QPCs)
connecting the dot to source and drain. A mobility $\mu \sim
300,000$~cm$^{2}$/Vs is obtained at $n \sim 1.8 \times
10^{11}$~cm$^{-2}$, corresponding to $\ell \sim 2.1~\mu$m. The
modulation-doped heterostructure has $\mu \sim 333,000$~cm$^{2}$/Vs
at $n \sim 2.4 \times 10^{11}$~cm$^{-2}$ giving $\ell \sim
2.7~\mu$m. The 2DEG-dopant separation is $20$~nm, as in
\cite{MarcusPRL92}, where comparable mobility is obtained. M1 is
defined by side-gates (Fig.~1(b)) negatively biased to $V_{SG} <
-0.23$~V, more negative $V_{SG}$ decreases the dot area and QPC
width. Further details on the undoped devices are available
elsewhere~\cite{EPAPS, KaneAPL93, ClarkeJAP06, SeeAPL10}.

\begin{figure}
\includegraphics[width = 6cm]{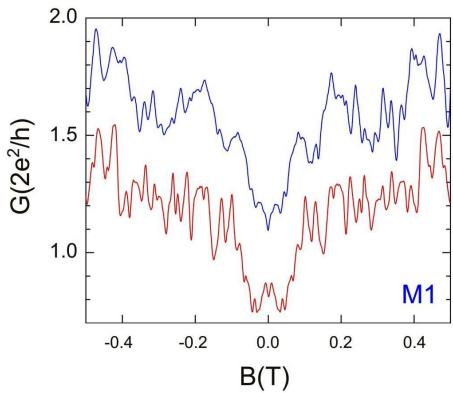}
\caption{Magnetoconductance $G(B)$ vs perpendicular magnetic field
$B$ for M1 with $V_{SG} = - 443.5$~mV before (blue) and after (red)
thermal cycling to $T = 300$~K. The red trace is vertically offset
by $-0.6$ for clarity.}
\end{figure}

Electrical measurements were performed in the dark at a temperature
$T \sim 230$~mK, with $G(B)$ obtained by standard four-terminal
lock-in techniques with the magnetic field $B$ perpendicular to the
2DEG. The low temperature $G(B)$ shows quantum interference
fluctuations~\cite{StonePRL85} that provide a
`magnetofingerprint'~\cite{FengPRL86} of the distribution of
electron trajectories through the dot. Note that in addition to dot
geometry, $G(B)$ is highly sensitive to the disorder
distribution~\cite{FengPRL86}. To isolate the effect small-angle
scattering has on transport, we examine the changes in $G(B)$
induced by thermal cycling of dots with and without modulation
doping. This relies on the well-known tendency for the disorder
potential to differ between cooldowns in modulation-doped
Al$_{x}$Ga$_{1-x}$As heterostructures with $x \gtrsim 0.22$ due to
the capture of excess electrons by deep donors known as DX centres
~\cite{MooneyJAP90, LongSST93}. In modulation-doped devices, this
leads to significant changes in $G(B)$ upon warming above $T \sim
150$~K despite the defined dot geometry remaining exactly the
same~\cite{BervenPRB94, ScannellarXiv11}, as shown in Fig.~2, where
we plot $G(B)$ for device M1 before and after thermal cycling to $T
= 300$~K. Many attempts were made to obtain reproducible $G(B)$
traces from M1 without success~\cite{EPAPS}.

\begin{figure}
\includegraphics[width = 8.5cm]{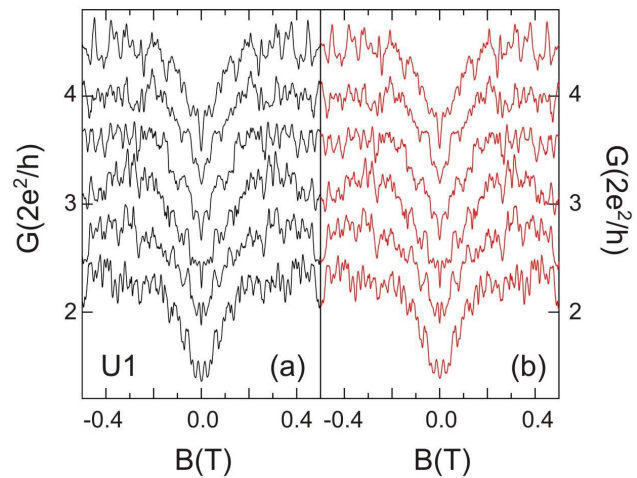}
\caption{$G(B)$ vs $B$ for U1, increasing $V_{TG}$ from $+ 930$~mV
(top) to $+ 955$~mV (bottom) in $5$~mV steps, (a) before (black) and
(b) after (red) thermal cycling to $T = 300$~K.}
\end{figure}

Figure~3(a/b) presents $G(B)$ data from the first and second
cooldowns of device U1, with side-by-side comparisons for six
different $V_{TG}$. {\it Remarkably, $G(B)$ is reproducible with
high-fidelity in U1 despite thermal cycling to room temperature, in
stark contrast to modulation-doped devices.} We observe this
behavior in other undoped devices and for repeated cooldowns of a
single undoped device~\cite{EPAPS}. We used a cross-correlation
analysis to quantify the extent of changes in $G(B)$ due to thermal
cycling, with the correlation $F$ normalized to give $F = 1$ ($F =
0$) for identical (randomly-related) traces~\cite{TaylorPRB97,
EPAPS}. We obtain $F = 0.94$ for the upper two U1 data traces in
Fig.~3(a/b) ($V_{TG} = +930$~mV), compared to a maximum of $F =
0.75$ for M1, mostly due to the similar $G(B)$
background~\cite{EPAPS}.

The fact that $G(B)$ is reproducible after thermal cycling for U1
but not for M1 demonstrates that the ionized dopant disorder
potential has a significant, perhaps even dominant, influence on the
electron dynamics. It immediately shows that small-angle scattering
cannot be ignored and that the commonly-held simplistic picture of
electrons following straight-line trajectories that undergo specular
reflections at the dot walls is unrealistic. The question is, how
does this affect our understanding of electron transport in quantum
dots? Two aspects are involved here: length scale and the effect on
experimental signatures of the dynamics.

\begin{figure}
\includegraphics[width = 8.5cm]{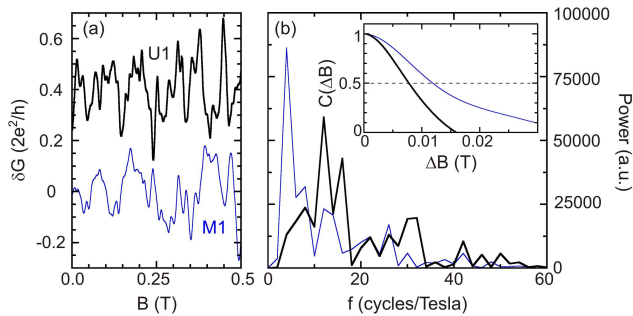}
\caption{(a) Extracted magnetoconductance fluctuations $\delta G$ vs
$B$ for U1 (uppermost left trace in Fig.~3 -- thick black line) and
M1 (uppermost trace in Fig.~2 -- thin blue line). The U1 data is
vertically offset by $+0.4$ for clarity. (b) Fourier power spectra
and (inset) normalized autocorrelation functions obtained for the
two traces in (a).}
\end{figure}

Dealing with length scale first; from a semiclassical view, ignoring
disorder, there is a wide distribution of lengths in the set of all
possible electron trajectories within the dot that intercept the
QPCs and thus contribute to $G(B)$~\cite{BarangerChaos93}. One can
naturally expect that the impact of small-angle scattering to
increase with trajectory length, being minimal for the shortest
trajectories. SGM studies of dots show clear evidence that shorter
ballistic trajectories survive the diffusive effect of small-angle
scattering~\cite{CrookPRL03, BurkePRL10}. This suggests that studies
focussed on the influence of short periodic orbits on $G(B)$
spectral content~\cite{BirdPRL99} may be robust, partly due to the
reduced impact of small-angle scattering for short trajectories, but
perhaps also because quantum interference may enable these orbits to
survive despite the diffusive effect of small-angle
scattering~\cite{HellerPRL84, BurkePRL10, ZozoulenkoPRB97}.
Regarding longer paths, the difficulty is that $G(B)$ reflects the
distribution of areas enclosed by possible trajectories, and this
does not directly map to the trajectory length distribution due to
flux-cancellation effects~\cite{BeenakkerPRB88}. Thus changes in
disorder potential will impact broadly across the spectrum of $G(B)$
fluctuations. One case where longer paths may be more robust to the
disorder potential are skipping orbits running along the dot walls
at moderate $B$~\cite{ChristenssonPRB98}. These may be reinforced by
the process described by B\"{u}ttiker (see Fig.~4 of
\cite{ButtikerPRB88}). The feasibility of this is evident in the SGM
studies by Aidala {\it et al}~\cite{AidalaNP07}.

To demonstrate that small-angle scattering has a tangible effect on
the statistics of $G(B)$ fluctuations, in Fig.~4(b) we present
Fourier power spectra and autocorrelation analyses of representative
data from U1 and M1 (Fig.~4(a)); we obtain qualitatively similar
results for other traces (see Fig.~S3 in \cite{EPAPS}). The
fluctuations are extracted by symmetrizing the data, and removing a
third-order polynomial fit as per \cite{MarcusPRL92}. The rms
amplitudes are similar ($0.0985$ and $0.0925 \times 2e^{2}/h$ for U1
and M1), but higher frequency fluctuations are clearly evident for
U1. This is borne out in Fig.~4(b), where the U1 spectra shows
enhanced power at higher frequencies relative to M1, confirmed by
the autocorrelation analysis~\cite{JalabertPRL90, MarcusPRL92} inset
to Fig.~4(b). The correlation $C(\Delta B) = \langle\delta
G(B)\delta G(B + \Delta B)\rangle$ drops more rapidly for U1,
consistent with richer structure in the fluctuations (the
correlation fields for U1 and M1 are $7.8$ and $11.9$~mT). An
interesting aspect of Fig.~4(b) is a distinct tendency towards a
higher frequency for the dominant peak in the U1 spectra; this may
point directly to the influence of disorder on transport given that
U1 and M1 have nominally identical geometry and differ in the
presence/absence of small-angle ionized dopant scattering.

Turning now to how small-angle scattering affects experimental
signatures of the electron dynamics, the most obvious is $G(B)$
itself. Berry {\it et al}~\cite{BerryPRB94} proposed that $G(B)$
could be directly used as a magnetofingerprint to detect the change
in electron dynamics induced by adding a narrow barrier to the
interior of a circular quantum dot. This required two separate
devices, and our findings show that an equivalent change in $G(B)$
would have resulted from room temperature thermal cycling of either
of these modulation-doped devices. The same problem exists for more
recent work also, e.g., \cite{MicolichPRB04}. Simple statistical
measures are also affected; for example, Fig.~3 of Marcus {\it et
al}~\cite{MarcusPRL92} presents power spectra for two separate
nominally identical device `chips', each containing one circular and
one stadium-shaped dot. If small-angle scattering from the disorder
potential was negligible, one would expect the power spectra for the
circular dot on each chip to be very similar, the same should hold
for the two stadia. Indeed, differences in spectra between stadium
and circle are no more or less substantial than those between two
identical dot geometries in Fig.~3 of \cite{MarcusPRL92}. This
suggests these spectral differences are not due to geometry alone
but also reflect differences in disorder potential (e.g. see Fig.~S4
of \cite{EPAPS}).

It is important to note that in both cases above we do not claim
that dot geometry has no effect on electron dynamics at all, only
that small-angle scattering masks its effect on $G(B)$ as an
experimental signature of dynamics. One approach that may overcome
the effect of small-angle scattering is that used by Chang {\it et
al}~\cite{ChangPRL94}, where a $6 \times 8$ array of nominally
identical dots was measured to average out the $G(B)$ fluctuations.
Our findings suggest this essentially constitutes an averaging of
the dot disorder potentials to expose the effect of the common
lithographic geometry. Although the link between zero-field
conductance peak lineshape and dynamics has been
questioned~\cite{BirdPRB95, AkisPRB99}, this dot array approach may
be useful for comparing other aspects of the measured $G(B)$ to
theoretical predictions, providing the array of dots is sufficiently
identical lithographically. Such studies would be aided considerably
by the undoped device architecture~\cite{SolomonEDL84, KaneAPL93,
SeeAPL10}. More complex spectral analyses involving significant
post-measurement averaging of $G(B)$ fluctuations may also be
helpful.

Although removal of modulation doping significantly reduces
small-angle scattering, we expect substantial disorder to remain,
e.g., background impurities, interface roughness. It is difficult to
comment further on this remnant disorder, but our data shows that it
is robust to thermal cycling. We suspect this remnant disorder will
prevent perfectly identical $G(B)$ from being obtained in separate,
nominally-identical undoped devices, obviating the truly ballistic
quantum dots envisioned theoretically~\cite{JalabertPRL90,
JensenNat95}, but the considerably improved thermal robustness of
our undoped architecture, along with the ability to reduce {\it
both} large- and small-angle scattering, makes it highly appealing
towards potential practical applications of ballistic transport
devices~\cite{GoodnickIEEE03}.

Our results also have more broad implications for our understanding
of disorder in mesoscopic devices, highlighting the limitations of
mobility as a metric for disorder. For example, some may argue that
HIGFETs failed to meet expectations because the highest mobilities
obtained fall short of those in modulation-doped structures. We
believe this belies the truth, because as pointed out earlier, the
mobility is heavily weighted towards large-angle
scattering~\cite{ColeridgePRB91}, and hence small-angle scattering
is more or less ignored considering mobility alone.  The fact that
small-angle scattering has a major effect on transport at length
scales much smaller than expected from the mobility (i.e., at the
$\sim 20$~nm scale rather than $\sim 2000$~nm) is evident in both
SGM studies~\cite{JuraNP07, AidalaNP07} and the lack of $G(B)$
reproducibility under thermal cycling in modulation-doped dots. The
importance of this broader picture of disorder is also evident in
very recent studies of the $5/2$ fractional quantum Hall
state~\cite{PanPRL11}, driven by interest in the $5/2$ state for
topological quantum computation, and the associated need for
ultra-low disorder heterostructures. From this perspective, we
suggest the $G(B)$ in quantum dots may be highly useful for
detecting changes in the small-angle scattering potential whilst
studying scattering mechanisms in high-mobility modulation-doped
heterostructures~\cite{LarocheAPL10}. Alternatively, shallow dots in
inverted undoped heterostructures~\cite{MeiravAPL88} might be used
to detect changes in charge environment at or above the
heterostructure surface; for example; using chemical treatments
between thermal cycles may aid the study of surface charge effects
on transport~\cite{MakAPL10}, or enable dots to be used for charge
detection more generally. Dopant reconfiguration is also a well
known source of charge noise in modulation-doped
devices~\cite{BuizertPRL08}; our undoped devices' thermal robustness
suggests they may offer a path to devices with significantly reduced
charge noise.

This work was funded by the Australian Research Council (ARC)
[DP0772946, LX0882222, FT0990285], Office of Naval Research
[N00014-07-0457], US Air Force [FA8650-05-1-5041], National Science
Foundation and Research Corporation, and performed  using the
Australian National Fabrication Facility. APM/ARH acknowledge ARC
Future/Professorial Fellowships.

\end{document}